\documentstyle[preprint,aps,graphicx,amssymb]{revtex}

\newcommand{\etal}{{\em et al.}}                
\tighten
\begin{document}
\draft
\title{Spectator Response to the Participant Blast\\
}
\author{L.~Shi, P.~Danielewicz
}
\address{
National Superconducting Cyclotron Laboratory and
 Department of Physics and Astronomy,\\
Michigan State University, East Lansing, MI 48824\\
}
\author{R.~Lacey
}
\address{
Department of Chemistry, State University of
New York at Stony Brook,  \\
Stony Brook, NY 11794-3400
}
\date{\today}
\maketitle
\begin{abstract}
The interplay between spectator and participant matter in heavy-ion
collisions
is investigated in the context of a microscopic transport
model. Transport simulations show that
flow patterns for the participant matter are strongly
influenced by the presence of the nearby
spectator matter.  However, the influence is mutual.
During the explosion of the participant zone, the
spectator matter acquires a transverse momentum $\langle P^x /A
\rangle$
that shows sensitivity to the nuclear incompressibility and to
the momentum dependence of the nuclear mean field (MF).
An observed change
in the net average momentum per nucleon, $\Delta |\langle {\bf P}/A
\rangle|$,
can be associated with the momentum dependence of the MF.
For a momentum-dependent MF and a low impact parameter in a
heavy
system, the spectators may emerge faster than in the initial
state, accelerated by the violent participant explosion.
The average
excitation energy and the mass of the spectators, in contrast
to the momentum, show little sensitivity to the
nuclear equation of state.

\end{abstract}
\pacs{PACS. 25.75.-q, 25.70.z, 25.75.Ld, 21.65.+f}

\section{INTRODUCTION}
\label{sec:intro}
One of the major goals of intermediate and high-energy heavy-ion
collision studies is the
determination of the nuclear equation of state (EOS).
The EOS constrains the nucleon-nucleon interactions in nuclear matter
\cite{Wiringa88} and is an important ingredient for a detailed
understanding
of the supernovae explosions~\cite{lie00}, the structure of
neutron stars~\cite{pra97},
and the evolution of the early Universe~\cite{boy01}.  While
the precise relationship
between the EOS and the nucleon-nucleon interaction remains
complicated~\cite{Wiringa88}, the
EOS continues to be an important ingredient for developing an
understanding for a range of
nuclear processes including many
heavy-ion reaction-phenomena~\cite{Bertsch88}.

Phenomenological parameterizations of the EOS are usually constrained
with the
properties of nuclear-matter at normal density $\rho_0$ and
diverge
at much higer densities that can be probed in energetic
heavy-ion reactions. An important
complication for heavy-ion collisions
results from the fact that the duration of the initial high-density
stage of the collision is very short compared to the time scale for the
whole reaction process.  E.g.,
in an 800~MeV/nucleon $b=5$~fm collision of
$^{124}$Sn + $^{124}$Sn,
the
high-density stage with a central density $\rho_c > 1.5 \,
\rho_0$ lasts about $13$~fm/c, while the elapsed time from the
initial impact to
the complete seperation of target and projectile is $\sim$
$40$~fm/c.  The~spectator properties  continue to develop
well beyond this time\cite{Wolter99}. Given the short duration of the
high-density stage, signals which carry information about the
high-density phase of the collision could be easily washed
out by other
signals generated at a later stage.  In consequence, reaction
simulations are
needed to provide guidance for the measurement of signals which not only
probe the high-density stage but survive through the entire
duration of the collision process.

The collective flow of participants has been observed for a
long time \cite{Doss86,Gustafsson88,Gutbrod90}.
The flow is believed to result from
early stage compression and an expansion
\cite{Reisdorf97,Pak96,Danielewicz92,Danielewicz95,Danielewicz98},
and can carry information
on the initial high-density phase. The relation between the
nuclear EOS and
the flow phenomena has been explored extensively in simulations
and a recent
example is the analysis of the transverse-momentum dependence
of
the elliptical flow \cite{Danielewicz00}.  The elliptical flow
is
shaped by an interplay of geometry and mean field and, when
gated by
the transverse momentum, reveals the momentum dependence of
mean field at supranormal densities.

The elliptical flow pattern of the participant matter is
affected by the presence of the cold
spectators \cite{Danielewicz98,Danielewicz00,Larionov00}, as
will be reiterated.
When nucleons are deccelerated in the participant region,
the longitudinal kinetic energy associated with the initial
colliding nuclei is converted
into the thermal and potential compression energy.
In a subsequent rapid
expansion or explosion, the collective transverse energy
develops~\cite{Danielewicz92,Danielewicz95,Reisdorf97,Pak96}
and many particles from the participant region
get emitted in the transverse directions.
The particles emitted towards the reaction plane can encounter
the cold spectator pieces and, hence, get redirected.  In
contrast, the particles emitted essentially perpendicular to
the reaction plane are largely unimpeded by the spectators.
Thus, for beam energies leading to a rapid expansion in the vicinity of
the spectators, elliptic flow directed out of the reaction plane
(squeeze-out) is expected.  This squeeze-out is related to the pace at
which the expansion develops, and is, therefore, related to
the~EOS.

On the other hand, since the spectators
serve to deflect particle emissions
toward the reaction plane, their properties may be significantly
modified.  This prompts
us to analyze the characteristics of the spectators resulting
from the collision process.  In one sense, the spectators can be viewed
as probes which were present
at the site of the nuclear explosion leading
to the rapid particle emission.  Thus, a careful study of their
characteristics could complement the results from the
elliptic flow and provide further information on the properties of
high-density
nuclear matter.

Long-time evolution of spectators has been studied in recent
past by Gaitanos {\em et al.}\cite{Wolter99}.
A comprehensive summary of experimental results for spectators produced
in reactions at different centralities has been presented by
Pochodzalla \cite{Pochodzalla00}. In particular,
universal features of spectator multifragmentation
has been well documented \cite{Pochodzalla00,Schuttauf96}.
The transverse momentum
change of the spectator during a semicentral collision, to be
addressed
here, was looked at in the past via emulsions by Bogdanov {\em
et al.} \cite{Sov91} (see also~\cite{che84}).
The systematics of the longitudinal momentum transfer to spectators
in energetic reactions induced by light projectiles can be found
in Ref.~\cite{mor89}.

In this paper, we present an analysis of the interplay between
participant
and spectator matter
during the violent stage of a heavy-ion reaction.
We examine the impact of the interplay on
the evolution of different physical characteristics
within the two reaction zones.
We investigate whether the spectator properties can be related
in any fashion to the high-density EOS.

The paper is organized as follows.  A brief description of our
transport model is given in Section\ \ref{sec:BUU}. The reaction process
and the participant-spectator interaction are described in Section\
\ref{sec:reaction}.
The sensitivity of the emerging spectator characteristics
to the nuclear EOS is investigated in Section\ \ref{sec:sensitivity}.
The paper is summarized in
Section\ \ref{sec:summary}.

\section{Transport model}
\label{sec:BUU}

In this work, we utilize a set of transport equations of the
Boltzmann-Uhlenbeck-Uehling (BUU) type to simulate central heavy-ion
reactions \cite{Bertsch88}.
Within the approach, the reacting system is represented by
quasiparticles with definite momentum and energy.  Those
quasiparticles move within the mean field (MF) produced by
other quasiparticles and undergo collisions that represent
fluctuations in the interaction with the medium beyond the mean
field.  With this, the phase-space
distributions of quasiparticles, $f_X \equiv f_x({\bf p},{\bf
r},t)$, follow the set of equations:
\begin{equation}
{\partial f_X \over \partial t}+{\partial \epsilon_X \over
\partial {\bf p}} {\partial f_X \over
 \partial {\bf r}} - {\partial \epsilon_X \over \partial {\bf
r}} {\partial f_X \over \partial {\bf p}}
=  {\cal K}^<_X(1 \mp f_X) - {\cal K}^>_X f_X~.
\label{eq:1}
\end{equation}
The l.h.s.\ in the above accounts for the changes in the
distribution caused by the the MF and the r.h.s.\ for
the changes in the distribution due to collisions.
The single-particle energies $\epsilon_X$ are variational
derivatives of the total system energy with respect to the
phase-space distribution.  The dependence of the total energy
on phase-space distributions yields in the equilibrium the
system EOS.

We carry out our calculations utilizing strongly
momentum-dependent
(MD) and momentum-independent (MI) MFs.  In the
calculations with the MI MFs, the energy density in
the system is
\begin{equation}
{e} = \sum_X g_X \int {d{\bf p} \over {( 2 \pi )^3}} \,
f_X({\bf p}) \sqrt{p^2+m^2_X(\rho_s)}
+\int^{\rho_s}_0 \, d\rho_s' \, U(\rho_s') -\rho_s \, U(\rho_s)
~ . \label{eq:2}
\end{equation}
where $m_X(\rho_s)=m_X+A_X \, U(\rho_s)$, $A_X$ is baryon
number and $\rho_s$ is scalar baryon density.
The single-particle energy is then
\begin{equation}
{\epsilon}_X(p, \rho_s) = \sqrt{p^2+m^2_X(\rho_s)} \, ,
\label{eq:3}
\end{equation}
and we parametrize the scalar field $U$ with
\begin{equation}
U(\rho_s)={-a \left(\rho_s / \rho_0 \right) +b
\left(\rho_s / \rho_0 \right)^{\nu} \over
{1+{\left({\rho_s \over 2.5
\, \rho_0}\right)}^{\nu-1}}} \, , \label{eq:4}
\end{equation}
where $\rho_0$ is the normal density
and $a$, $b$ and $\nu$
are parameters.
In the calculations with MD MFs, the energy density is
represented in the local baryon frame as
\begin{equation}
{e}=\sum_X g_X \int {d{\bf p}\over {(2 \pi)^3}}f_X({\bf p})
\left( m_X+\int^p_0 dp' v^*_X(p',\rho)\right)+\int^{\rho}_0 d\rho'
U(\rho')~ .
\label{eq:5}
\end{equation}
The adopted form for
$U(\rho)$ is the same as in (\ref{eq:4}) and the local particle
velocity is parameterized with
\begin{equation}
v^*_X(p,\rho)= {p \over \sqrt{p^2+{m^2_X \over \left(1+c{m_N\over
m_X}{A_X(\rho/\rho_0) \over (1+\lambda p^2/m^2_X)^2} \right)^2 } }}~.
\label{eq:6}
\end{equation}
With the above, the single-particle energy in the local frame
is:
\begin{equation}
{\epsilon}_X(p,\rho) = m_X +\int^{p}_0dp' v^*_X+A_X \left[ \rho
\left< \int^{p_1}_0 dp'
{\partial v^* \over \partial \rho}\right>+U(\rho)\right]~.
\label{eq:7}
\end{equation}

The parameters for the MFs are given in Table \ref{tabmo}.
Other details of our calculations, such as concerning the
initialization of
nuclei and the lattice Hamiltonian method for integrating the
equations, can be found in Ref.~\cite{Danielewicz00}.
For nucleon-nucleon collisions, a moderated elastic in-medium
cross section is utilized through out this paper
(same as Ref.~\cite{Danielewicz00}), unless otherwise indicated.

\section{Spectators and Participants}
\label{sec:reaction}

The
important concept
of spectators and participants in collisions
was
first introduced by Bowman \etal\cite{bow73} and later
employed for the description of a wide-angle energetic particle
emission by Westfall \etal\cite{wes76}.
The two nuclei
slamming against one other can be viewed as producing
cylindrical cuts through
each other.  The swept-out nucleons or
participants
(from projectile and target) undergo a violent collision
process.
The remnants of the projectile and target, the
spectators, continue in the meantime with largely
undisturbed velocities, and are much less affected by the
collision process than the participant nucleons.
On one hand, this picture is supported by
features
of the data and, on the other, by dynamic simulations\cite
{Gosset77,Danielewicz92,Danielewicz95,Wolter99}.
During the violent stage of a reaction the spectators can
influence the behavior of participant matter.
Specifically,
the spectators can inhibit
the collective transverse expansion of the decompressing
participant matter
and effectively shadow particle emission directed toward the
reaction plane.  A consequence of this shadowing is the
observation of preferential squeeze-out of particles
perpendicular to the reaction plane as illustrated below.

Figure~\ref{v2t} summarizes the results obtained from simulations of
$^{197}$Au + $^{197}$Au collisions at a beam energy
$T_{lab}= 1$ GeV/nucleon and an impact parameter $b=8$ fm.
Unless indicated otherwise, a hard momentum-dependent (HM) EOS
(cf.\ Table~\ref{tabmo}) was used.
Figure~{\ref{v2t}(a)} shows the time evolution of the density of both
participant
and spectator matter.  The solid and dashed
curves show, respectively, the
baryon density $\rho_c$ at the center of the collision system
(participants) and
the baryon density in the local frame at the geometric
center of the spectator matter $\rho_{spec}$.
Here, the operational definition of spectator matter is that the
magnitude of the local
longitudinal velocity exceeds half of the velocity in the initial state
and that the local density exceeds one tenth of the normal density.  The
solid line in Fig.~{\ref{v2t}(a)} clearly illustrates
the rapid density build-up (for $t \leq 5$~fm/c) followed by
expansion of
the participant matter. The dashed-line also points to a weak
compression of the spectator
 matter during the expansion phase of the participants. The latter
observation is consistent
with the expected delay associated with the time it takes a compression
wave to reach
the center of the spectator matter, starting from an edge.

Figure~{\ref{v2t}(b)} shows the time evolution of the elliptic flow parameter
$v_2$ for
all mid-rapidity particles.  The parameter is defined as
\begin{equation}
v_2= \langle \cos{(2\phi)} \rangle \, ,
\end{equation}
where $\phi$ is the azimuthal angle in the X-Y plane perpendicular
to the beam axis Z; the X-Z plane defines the reaction plane.
The value of $v_2$ conveys information about the pattern of particle
emission from the central participant region.  The
hot participant region has an initial elliptical shape in the X-Y
plane due to the overlap geometry. Since the long and short
axes of the ellipse
point in the Y-direction and in the X-direction,
respectively, the matter starts
out with stronger MF and pressure gradients in the X-direction.
Given the shape of the emission source and
the gradients, the matter is first expected to develop
a stronger
expansion in the X direction and, hence, to give rise to
positive values of $v_2$.
If the spectators are nearby during the expansion phase, they can serve
to stall
the expansion in the X direction and a compression wave
then develops within the
spectator matter (cf. Fig.~{\ref{v2t}(a)}).  The resulting dominant
expansion of
participant matter in the Y direction gives rise to negative
values of $v_2$.  Figure~{\ref{v2t}(b)}  indicates that this preferential
out-of-plane emission
pattern begins after $\sim 7$ fm/c.
The time correlation between the change in sign of $v_2$ and the
decrease in the magnitude of the central density should be noted in the
figure.
The central density of participant matter $\rho_{c}$ begins to
drop at about $7$ fm/c and the most rapid declines ends
at $\sim 16$ fm/c; during this time the elliptical flow drops from its
maximum positive
value to its maximum negative value.

A comparion of Figs.~{\ref{v2t}(b)} and {\ref{v2t}(c)}
illustrates the important role
of the MF in shaping the elliptical flow magnitude. Figure~{\ref{v2t}(c)}
shows the
time dependence of $v_2$ obtained when the calculations are performed
without
the inclusion of a MF (cascade mode). In contrast to the
evolution with the mean field (cf. Figs.~{\ref{v2t}(b)}) where $v_2$ first
achieves
significant positive and then negative values, Fig.~{\ref{v2t}(c)}
indicates $v_2$
values which stay close to zero over the entire time evolution of the
system.
This trend is related to the fact that in the cascade model the
transverse expansion is
slow compared to the time duration for which the spectators are in close
proximity to the participant matter, or compared to the time
required for longitudinal
motion to thin the matter. The important role of the MF for the
generation of
elliptic flow and the sensitivity  of this flow to the EOS
has been stressed~\cite{Danielewicz98,Danielewicz00,Larionov00}.

The temporal difference of $v_2$ for
all midrapidity particles in the system, and for those particles that
have left
the system can be observed by comparing Figs.~{\ref{v2t}(b)}
and {\ref{v2t}(d)}.
Figure~{\ref{v2t}(b)} shows the change in $v_2$ with
time as discussed above. On the other hand, Fig.~{\ref{v2t}(d)} indicates
little or
no change of $v_2$ (over time) for midrapidity particles that
have left the system.
That is, out-of-plane emission is favored (negative $v_2$) for all
emission times.  Figures~{\ref{v2t}(b)} and {\ref{v2t}(d)} also
show $v_2$ as a function of time for
particles with transverse momentum $p_t > 0.55$~GeV/c; these
panels indicate that faster
particles are more sensitive to the obstructions as well as to
any directionality in the collective motion.

The analyses of elliptical flow and related works have
established connections
between features of the participant matter resulting from the
participant-spectator
interaction and the nuclear EOS
\cite{Danielewicz92,Reisdorf97,Pak96,Danielewicz00,Danielewicz98,Larionov00}.
On the other hand, it is not know whether the same interaction
(during the violent stage of a
reaction) leaves any lasting effects in the spectators
that could be related to EOS.  Extensive studies of the statistical
behavior of spectator matter have been carried
out\cite{Pochodzalla00,Schuttauf96}
for time scales which are long compared to the collision time. Such
studies
do not address the dynamical impact of the violent
reaction stage on spectators. During the violent stage of a collision,
the spectators remain close to the participant matter, so they might
serve as
a good sensor for the explosion.
Thus, we proceed to take a closer look at the changes which may occur in
the
spectator matter following their interaction with the participants. In
addition
we investigate whether or not such changes have a connection to the EOS.

\begin{figure}
\centerline{\includegraphics[angle=90,
width=.90\linewidth]{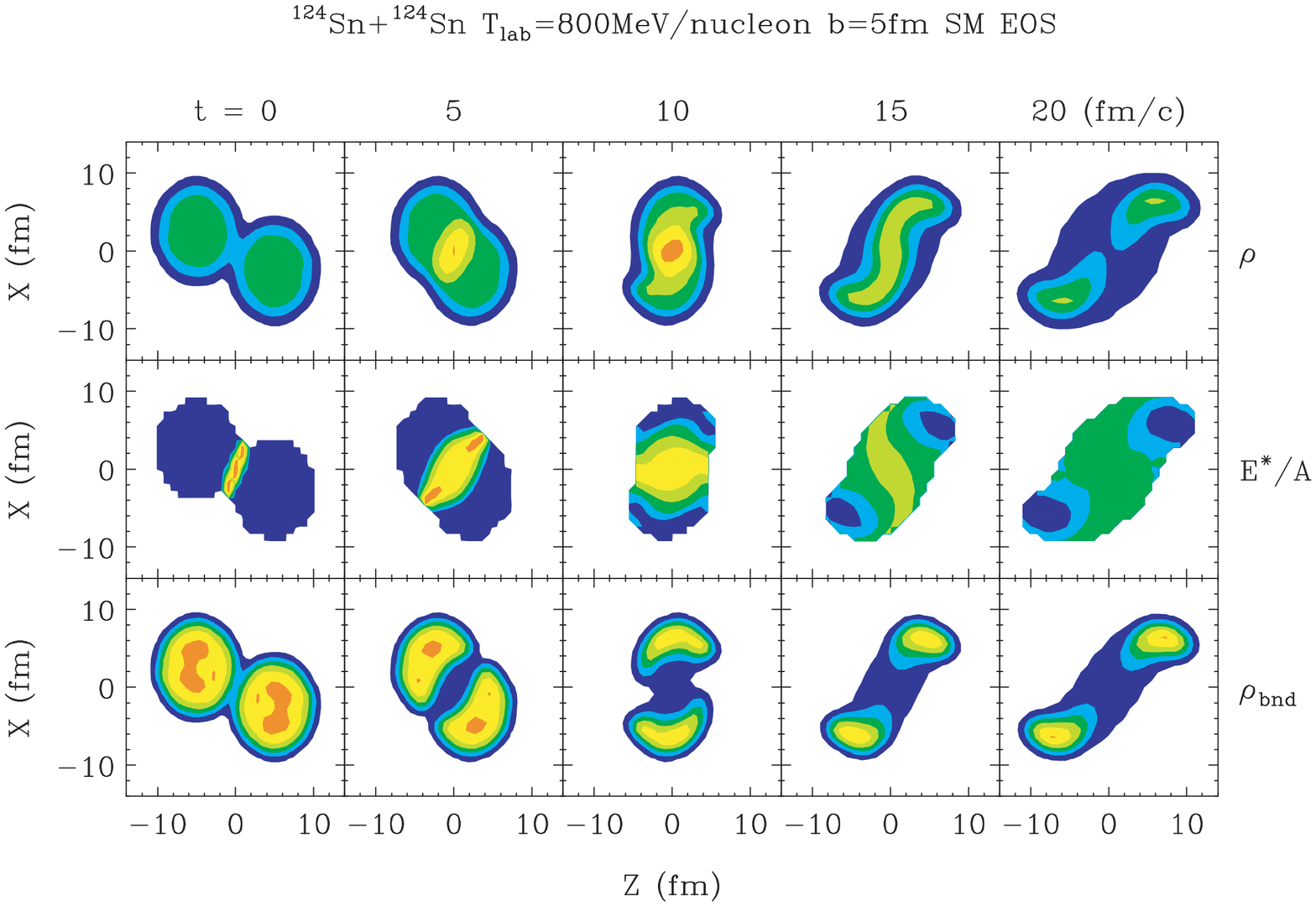}} 
\vspace*{.3in}
\caption{
Contour plots of the system-frame baryon
density $\rho$ (top row), local excitation
energy $E^*/A$ (middle row), and of the density
of bound baryons $\rho_{\rm bnd}$ (bottom row), in the
$^{124}$Sn + $^{124}$Sn reaction at
$T_{lab}=800$ MeV/nucleon and $b=5$ fm, at times
$t = 0$, 5, 10, 15 and 20~fm/c (columns from left to
right).  The calculations have been carriered out employing the
soft momentum-dependent EOS.
The contour lines for the densities correspond to the
values, relative to the normal density, of $\rho$ from
 0.1 to 2.1 with increment of 0.4.
The contour lines for $\rho_{\rm bnd}$ are from
0.1 to 1.1 with increment of 0.2.
The contour lines for the excitation energy correspond to the
values of
$E^*/A$ at 5, 20, 40, 80, 120, 160MeV.
 For statistical reasons, contour plots for the energy
have been suppressed for the baryon densities
$\rho < 0.1 \ \rho_0$.  Note, regarding the excitation energy, that the
interior of the participant region is hot while the interior of
the spectator matter is cold.
}
\label{contour}
\end{figure}

Figure~\ref{contour} shows contour plots of different
quantities within the reaction plane now from
$^{124}$Sn + $^{124}$Sn reaction simulations at the beam energy
of $T_{\rm lab}=800$ MeV/nucleon, at the impact
parameter $b=5$ fm, carried out with a soft momentum-dependent
(SM) mean field.
The columns from left to right represent the reaction at 5~fm/c
time increments.
The top and middle rows show the
baryon density in the system frame $\rho$ and the
local excitation energy $E^*/A$, respectively. The bottom row
shows the density $\rho_{\rm bnd}$ of baryons that are
bound in their local frame ($\epsilon_X < m_X$).  As may be expected,
the
excitation energies reach rather high values in the participant region
but remain low within the spectator region throughout the
violent stage of the reaction.  Most of the particles in the the
participant
region are found to be unbound i.e.\ $\rho_{\rm bnd}$ is low.
On the other
hand, most of the particles within the spectator region are bound.
They move with velocities that are close to each other, and this
keeps $\rho_{\rm bnd}$ sizeable throughout the violent collision
stage.

\begin{figure}
\centerline{\includegraphics[angle=180,
width=.70\linewidth]{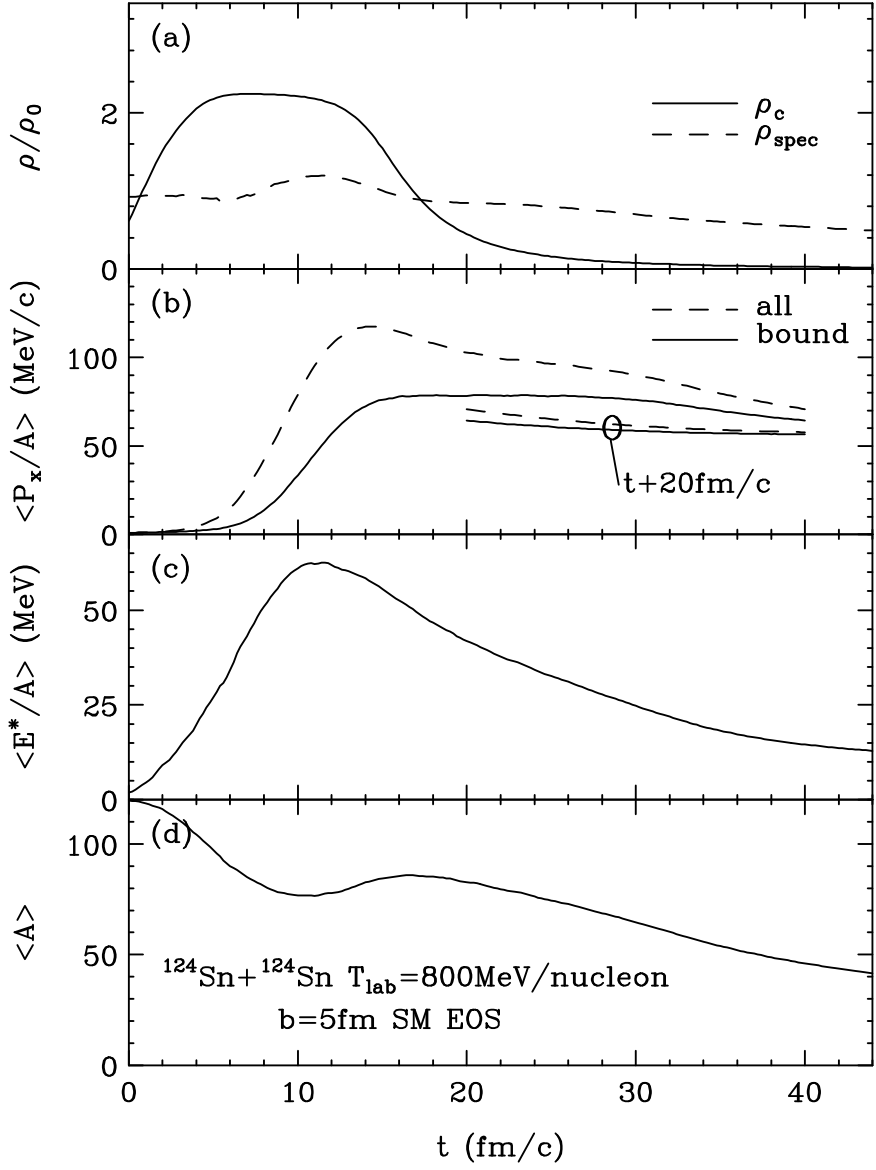}} 
\vspace*{.2in}
\caption{
Evolution of selected quantities in the $^{124}$Sn +
$^{124}$Sn reaction at 800 MeV/nucleon and $b=$ 5 fm, from a
calculation with a soft momentum-dependent EOS.
The panel (a) shows the density at the center of a
spectator~$\rho_{\rm spec}$ (dashed line) together with the
density at the
system center $\rho_c$ (solid line).  The panel (b) shows
the average in-plane transverse momentum per nucleon of the
spectator $\langle
P^X/A \rangle$ calculated using all spectator particles (solid
line) and using only bound spectator particles (dashed line).
Two extra lines in the panel show evolution of the momenta past
the 40~fm/c of the abscissa.
The panels (c) and (d) show, respectively,
the spectator
excitation energy per
nucleon $\langle E^*/A \rangle$ and the mass number $\langle A
\rangle$ from all spectator particles. }
\label{rhoc}
\end{figure}

Figure \ref{rhoc} provides next a detailed time development of
the selected
quantities in the 800 MeV/nucleon $^{124}$Sn + $^{124}$Sn
system for which the contour plots were given.
Figure~{\ref{rhoc}(a)} displays the evolution of baryon density at
the system center, $\rho_c$, and of baryon density at the
center of the spectator region, $\rho_{\rm spec}$.
The high-density stage for the
participant matter in Fig.~{\ref{rhoc}(a)}, characterized by $\rho_c
> \rho_0$, lasts over a time that
is short in comparison to the time needed for a clear
separation of the target and projectile spectators from the
participant zone, cf.\ Fig.~\ref{contour}.  To observe
a~stabilization of the spectator properties we needed to follow
the particular reaction up to $\sim 60$~fm/c.  Longer-term
studies of the spectator development have been carried out
within the BUU approach~\cite{Wolter99}.  However, as the
spectators
approach equilibrium, they might be described in terms
of the statistical decay method owning at this
stage advantages over the BUU equation.

Figure~{\ref{rhoc}(b)} shows the average
transverse momentum per nucleon of the spectator, in
the reaction plane, as a function of time.  In calculating the
average, we can include all spectator particles as
specified before (dashed line in the panel) or the subset
of particles that are bound in the local frames (solid line).
The averages, obviously, approach the same asymptotic value
over time, but the approach is faster for the bound-particle
average.
Note that the extra lines in Fig.~{\ref{rhoc}(b)} represent the
evolution of the average momenta past the 40~fm/c of the
abscissa.
Calculated in either manner,
the spectator average momentum $\langle P^X/A
\rangle$ reaches its magnitude during the high-density
stage in the participant matter and only somewhat
reduces to
stable during the expansion that follows.  This suggests
that the spectators can, indeed, provide information on the
high-density stage of the collision.

Figure~{\ref{rhoc}(c)} shows the
average excitation energy per nucleon $\langle E^*/A \rangle$
of the spectator
as a function a time.  Within the studied time interval, the
excitation energy rapidly rises and decreases and then
changes at a slower pace.  During the violent reaction stage,
some
particles traversing from the participant into the spectator
matter contribute to the excitation of the spectator matter.
As time progresses, some of those particles will travel through
the matter and leave the spectators.  Some other will degrade
their energy within the spectator frame.  (Note: we
consistently continue with definition where the spectator
matter is that for which the c.m.\ local velocity is larger
than half the beam velocity and the local density exceeds the
tenth of normal.)

Figure~{\ref{rhoc}(d)} shows the mass number of
a spectator region as
a function of time.  The spectator mass number decreases
rapidly as particles dive into the participant region and then
the mass recovers somewhat, around the time of 20~fm/c, as some
particles
get through the opposite moving corona matter and join the bulk
of the spectator matter moving along the beam direction.
Later, a~gradual deexcitation slowly reduces the spectator
mass.

We have demonstrated in this section the interplay between the
participants and spectators.  We have shown how the elliptic
flow is generated due to that interplay and how the interplay
affects the spectator characteristics.  In the next section we
explore the sensitivity of spectator characteristics to the
EOS for the nuclear matter in collision.

\section{Spectator Sensitivity to the Nuclear Equation of
State}
\label{sec:sensitivity}

In the light that the changes of the spectator
properties
could probe the compression and explosion of the participant
matter, we follow the reaction simulations till a clear
separation of the spectators from the
participant matter and a stabilization of the spectator
characteristics.  We explore the sensitivity of the emerging
spectator properties to different assumptions on the nuclear
EOS.  The results could serve to initialize statistical
decay calculations for a complete description of a reaction.

In the following, we shall present a sample of our spectator
investigations, within the $^{124}$Sn + $^{124}$Sn system
in the beam energy range of 250 MeV/nucleon to 1
GeV/nucleon at impact parameters $b = 5-7$ fm.  We shall also
quote results from $^{197}$Au + $^{197}$Au at 1 GeV/nucleon.
We utilized
four different EOS explored in the past, of which the
parameters are given in Table~\ref{tabmo}.
We concentrated on the
quantities
that could be experimentally determined for the spectator,
and thus the average transverse momentum per nucleon $\langle
P^X/A \rangle$, the change in the average c.m.\ momentum per
nucleon $\Delta | \langle {\bf P} /A \rangle | $, the average
excitation energy per nucleon $\langle E^*/A \rangle$, and the
average mass $\langle A \rangle$ following the violent stage of
the reaction.  The change in the average c.m.\ momentum
is $\Delta | \langle {\bf P} /A \rangle | = \sqrt{ \langle
P^X /A \rangle ^2 +  \langle P^Z /A \rangle
^2} - (P/A)_i$.

\begin{figure}
\centerline{\includegraphics[angle=180,
width=.70\linewidth]{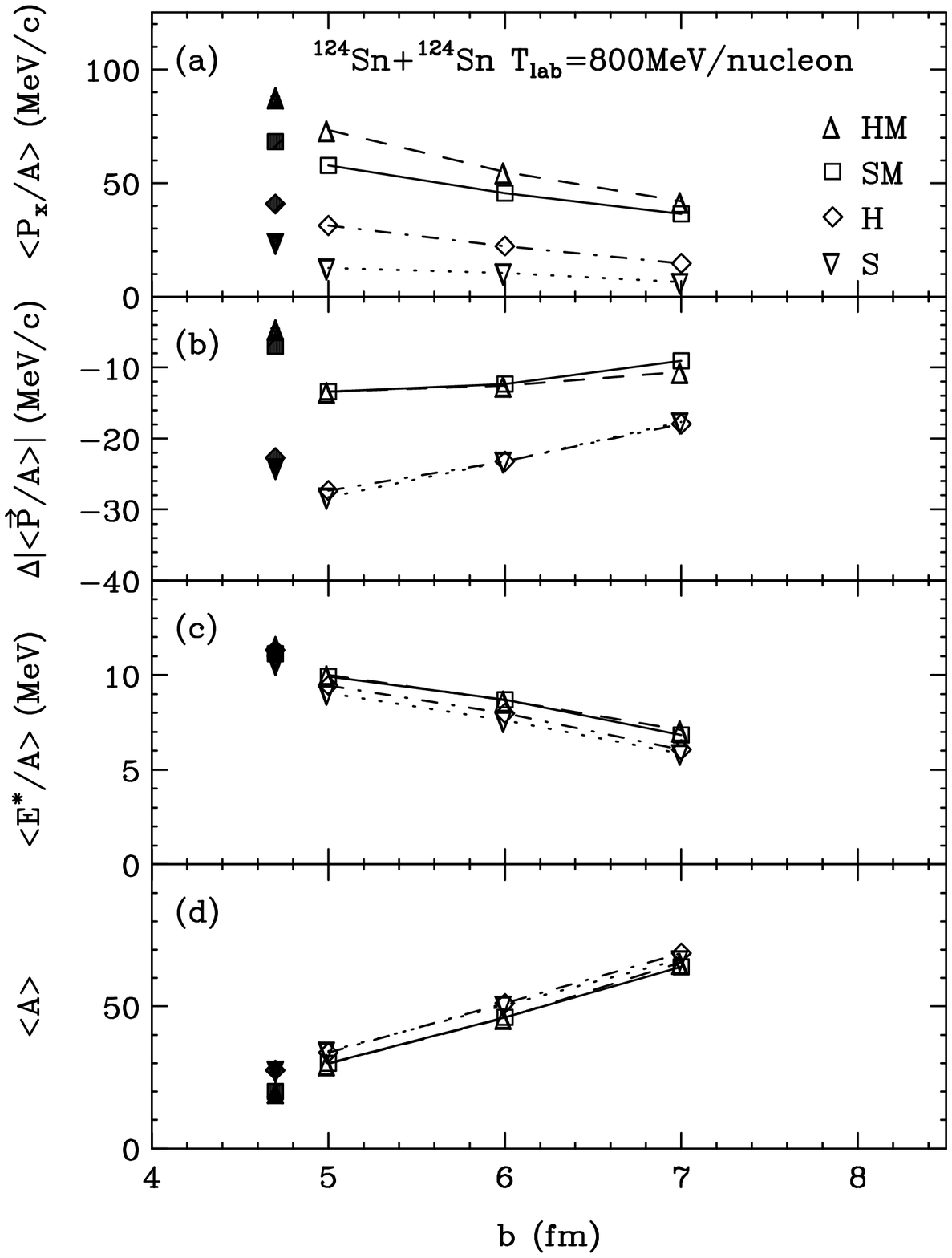}} 
\vspace*{.2in}
\caption{
Spectator properties in the 800 MeV/nucleon $^{124}$Sn +
$^{124}$Sn collisions, as a function of the impact parameter,
for four representative EOS: hard momentum-dependent (HM),
soft momentum-dependent (SM), hard momentum-independent (H) and
soft momentum-independent (S).
The panel (a) shows the average in-plane transverse momentum
of the spectator per nucleon $\langle P^X/A \rangle$.
The panel (b) shows the change in the average net c.m.\
momentum
per nucleon $\Delta |\langle {\bf P}/A \rangle|$.  The panel
(c)
shows the average excitation energy per nucleon $\langle E^*/A
\rangle$, and,
finally, (d) shows the average spectator mass
$\langle A \rangle$.
Open symbols represent results obtained with reduced in-medium
nucleon-nucleon cross sections; filled symbols represent results
obtained at $b=5$~fm with free cross sections.
}
\label{pxb}
\end{figure}

\begin{figure}
\centerline{\includegraphics[angle=180,
width=.70\linewidth]{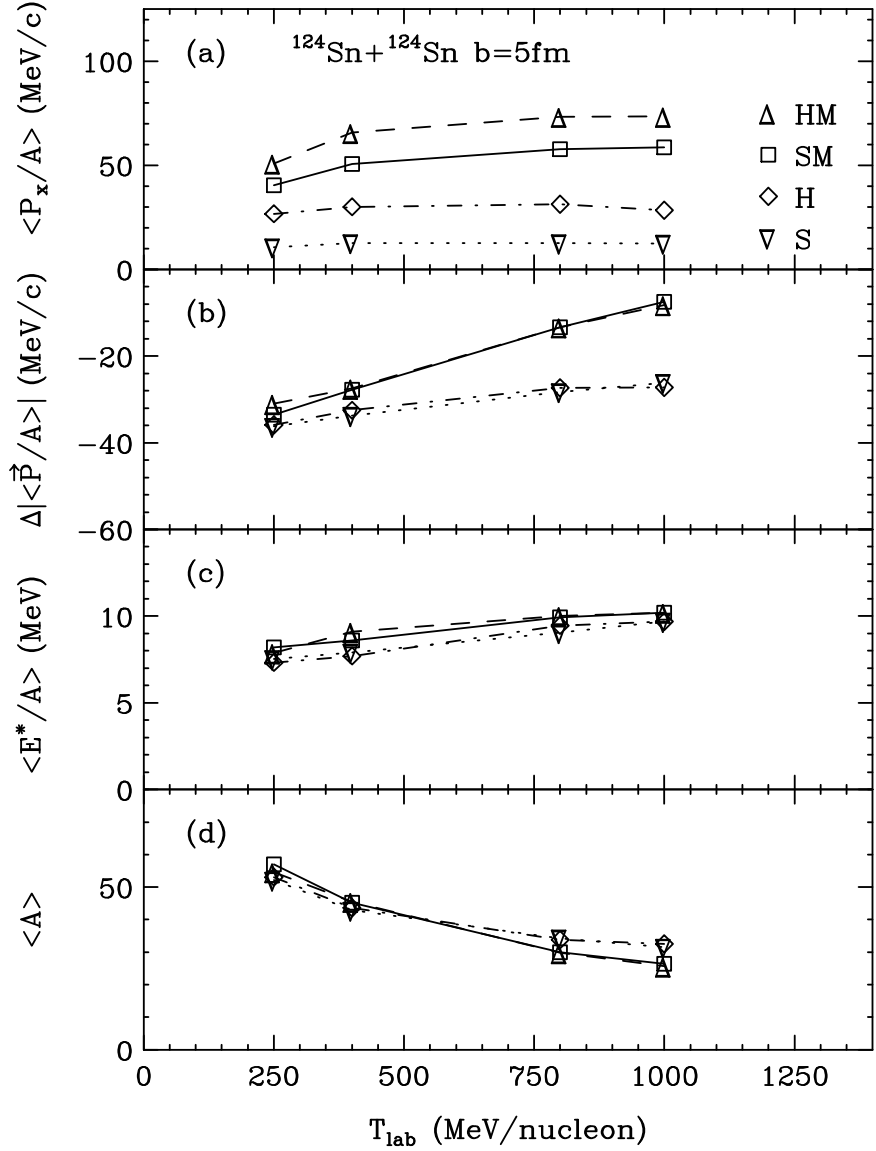}} 
\vspace*{.3in}
\caption{
Spectator properties in the $^{124}$Sn +
$^{124}$Sn collisions at $b=5$~fm, as a function of the beam
energy,
for four representative EOS: hard momentum-dependent (HM),
soft momentum-dependent (SM), hard momentum-independent (H) and
soft momentum-independent (S).
The panel (a) shows the average in-plane transverse momentum
of the spectator per nucleon $\langle P^X/A \rangle$.
The panel (b) shows the change in the average net c.m.\
momentum
per nucleon $\Delta |\langle {\bf P}/A \rangle|$.  The panel
(c)
shows the average excitation energy per nucleon $\langle E^*/A
\rangle$. Finally, the panel (d) shows the average spectator
mass $\langle A \rangle$.
}
\label{pxe}
\end{figure}

The above mentioned quantities, towards the end of our
simulations, are shown as a function of the impact parameter at
$T_{\rm lab} = 800$~MeV/nucleon in Fig.~\ref{pxb}, by open symbols,
and as a function of the beam energy
at $b=5$~fm in Fig.~\ref{pxe}, respectively.  The resulting spectator
$\langle P^X/A \rangle$ exhibits a clear sensitivity to the
stiffness of EOS.  We can see in both figures that a stiffer
EOS results in a stronger sidewards push to the spectator.
However, even more prominent is the sensitivity to the momentum
dependence of the mean field.  A strong momentum dependence
results in a stronger push to the spectator than the lack
of such dependence.  Recall that the interplay between the spectator and
the participant matter generates also the elliptic flow for the
participant matter and it was possible to exploit the latter in
the determination of the mean-field momentum-dependence at
supranormal densities
\cite{Danielewicz00,Larionov00}.

The final momentum of the spectator reflects the momentum
exchanges with the participant zone throughout the reaction.
Intially, the nucleons from the opposing nucleus move nearly
exclusively
along the beam axis relative to the spectators.  As
equilibration progresses, the momenta in the participant zone
acquire a level of randomness.  Random exchanges of momentum
between spectators and participants generally would drive the
spectator momentum towards the average for the system, i.e.\
zero.  However, the participant nucleons reach the spectators
moving away from the system center, coming with momentum
directed on the average outward, delivering an outward push to
the spectator pieces.

The order of magnitude for the transverse push may be obtained
from a simple estimate.  Thus, in Sn + Sn at 800 MeV/nucleon,
taking for the pressure in the
compressed region from the nonrelativistic ideal-gas estimate,
\begin{equation}
p \simeq \rho \, {2 \over 3} \, {T_{\rm lab}/4 A} \,
\end{equation}
with $\rho \sim 2 \, \rho_0 \sim 2/6$ fm$^{-3}$, we get $p \sim
40$ MeV/fm$^{3}$.
The size of the high-density region in the X-Z plane for Sn +
Sn at medium $b$ is $\sim 4$ fm, cf.\ Fig.~\ref{contour}.
The push to the spectator is then of the order of
\begin{equation}
 P^x \approx p \, S \, \Delta t \, ,
\end{equation}
where $S$ is the transverse area pushed by the participant
matter and $\Delta t$ is the duration of the push.  With $S =
\pi \, R^2/4 \sim 13$ fm$^2$ and $\Delta t \sim 5$~fm/c, cf.\
Figs.\ \ref{contour} and \ref{rhoc}, we get
\begin{equation}
{P^x \over A} = {40 \, \mbox{MeV/fm}^3 \, 13 \,
\mbox{fm}^2 \, 5 \, \mbox{fm/c} \over 50} \simeq 50 \,
{\mbox{MeV} \over \mbox{c}} \, ,
\end{equation}
This appears to agree as to the general magnitude with
what is found in the simulations.  When the impact parameter
increases, the fireball pressure decreases while the
spectator mass increases.  Thus, the momentum
per nucleon decreases.  With regard to the beam energy
variation in the simulations,
at low energies the pressure in the
fireball region drops, resulting in smaller push to the
spectators, with some
compensation coming from a longer time for the spectators in
the reaction zone and a longer fireball lifetime.  With the
rise in the beam energy from the low energy end,
the rise in the transverse fireball pressure is
moderated by the pion production and an increasing
transparency.  The spectator time in the vicinity of the
explosion continuously drops
resulting in a level of saturation in the spectator momentum
per nucleon.

With regard to the changes in the magnitude of the c.m.\
momentum per nucleon $\Delta | \langle {\bf P} /A \rangle | $,
we see in Figs.~\ref{pxb} and \ref{pxe} that the results for MD
MFs
significantly differ from the results for MI MFs for Sn + Sn,
with the latter MFs giving more momentum loss.  The spectator
mass and excitation energy, in contrast to the momentum, are
rather insensitive to the MF in the present system.

While the results discussed till now have been obtained with
reduced in-medium nucleon-nucleon
cross sections \cite{Danielewicz00}, we
also carried out calculations with the free nucleon-nucleon cross
sections. The latter calculations for the same system at
$T_{\rm lab} = 800$~MeV/nucleon $b=5$~fm, are represented by
filled symbol in Fig.~\ref{pxb}. With free cross sections, the remnant
masses are a bit lower, the excitation energies are higher, and so is
the transverse push. The transverse push is more sensitive to the change in
the EOS, than to the change in cross section, as evident in the figure.
Contrary to what one might naively expect, less momentum per nucleon
is lost in the free cross section case. We will come back to the
last issue later.

\begin{figure}
\centerline{\includegraphics[angle=180,
width=.70\linewidth]{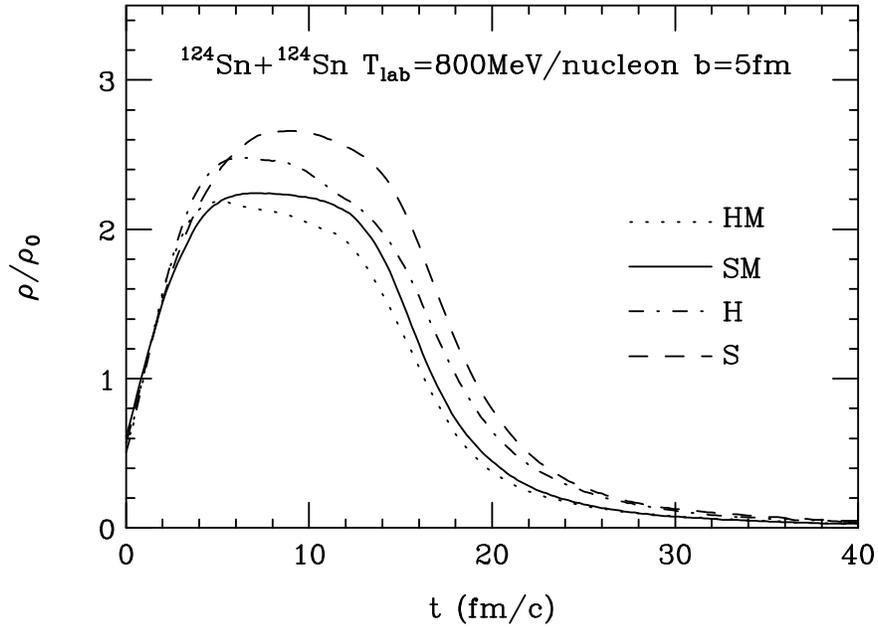}} 
\vspace*{.3in}
\caption{
Baryon density as a function of time at the center of the
$^{124}$Sn + $^{124}$Sn system at
$T_{\rm lab}=800$ MeV/nucleon and $b=5$ fm, for different MFs.
}
\label{rhot}
\end{figure}

In investigating the differences in results for the different
EOS, we obviously looked at the details in the time development
of the systems for the different EOS.  Figure~\ref{rhot} shows
the central participant density as a function of time.  For the
hard momentum-independent EOS
a maximal density is reached earlier and the expansion
sets faster than for the soft momentum-independent EOS.
The S EOS allows for a higher
compression than the H EOS.  An MD EOS allows for a lower
compression than a corresponding MI EOS.  Moreover, the
expansion develops earlier for an MD EOS than a corresponding
MI EOS.  Evidently, the momentum dependence plays a similar
role to the stiffness of nuclear matter; it renders the
matter less compressible in a dynamic situation.

\begin{figure}
\centerline{\includegraphics[angle=180,
width=.70\linewidth]{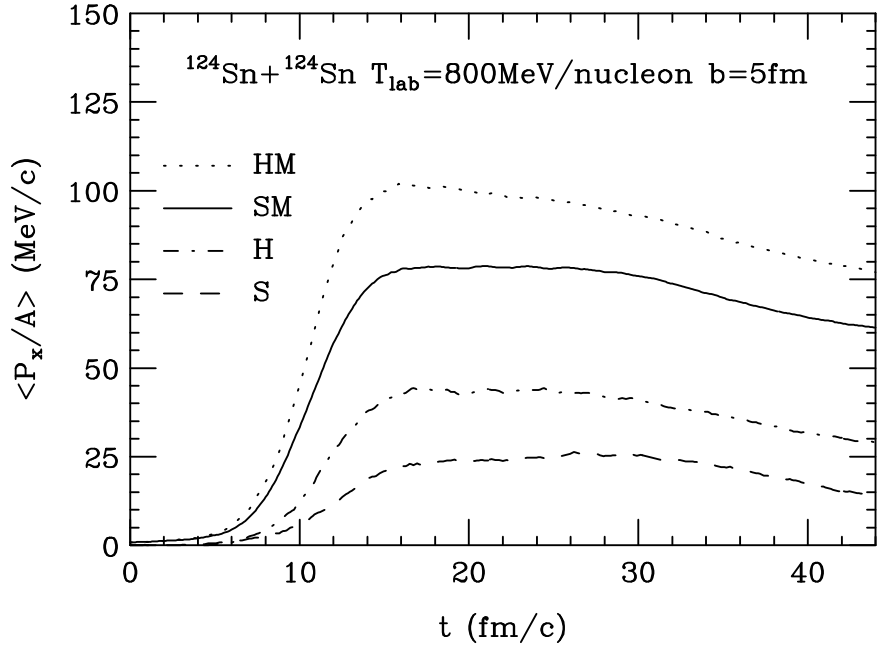}} 
\vspace*{.3in}
\caption{
Average in-plane transverse momentum per nucleon of a spectator
in $b=5$ fm
$^{124}$Sn + $^{124}$Sn collisions at
$T_{\rm lab}=800$ MeV/nucleon, as a function of time, for
different EOS.
}
\label{pxt}
\end{figure}

\begin{figure}
\centerline{\includegraphics[angle=180,
width=.70\linewidth]{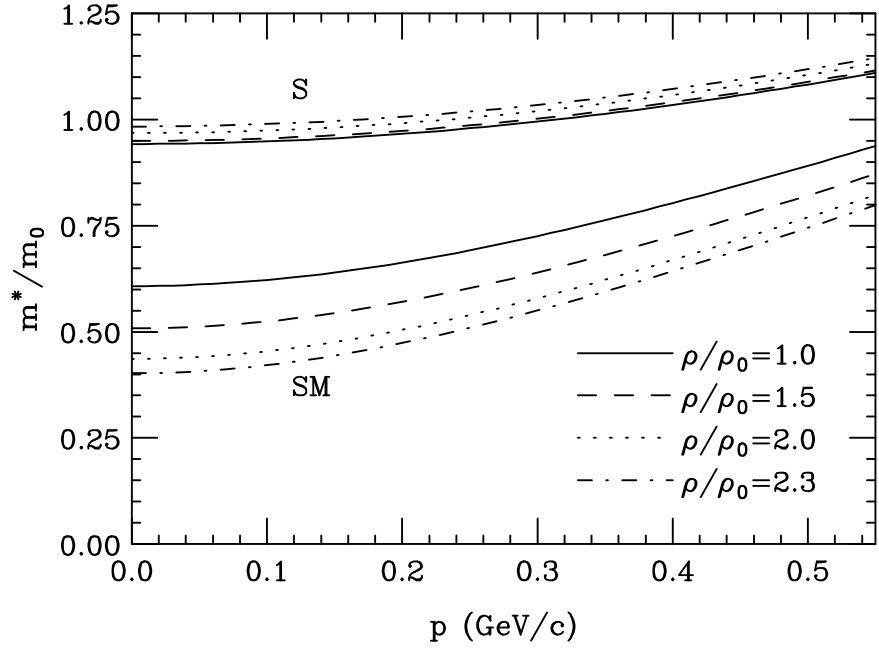}} 
\vspace*{.3in}
\caption{
Landau effective mass $m^* = p/v$, in units of free nucleon
mass,
as a function of momentum at several densities in cold nuclear
matter for S and SM MFs.
}
\label{eff}
\end{figure}

Figure \ref{pxt} shows next the spectator transverse momentum
in the X-direction as a function of time.  As we have already
pointed out before, the spectator transverse momentum per
nucleon rises within a relatively short time interval.
The rise starts about at the time when the maximal density is
reached at the participant center; the rise stops due to
combined effects of the spectator passing by and of the
dilution of the participant zone.  While there are up to 2~fm/c
differences in the start and end of the rise interval in
Fig.~\ref{pxt}, it is
apparent that the differences in the final
$\langle
P^X/A \rangle$
 must be due to the
differences
in magnitude of the transverse pressure (transverse momentum
flow) for the different EOS and not in the duration of the
rise.  In fact, the slopes of the
dependence of transverse momentum on time differ considerably
more than do the final transverse momenta.  A faster dilution for
the more incompressible EOS shuts off the momentum rise sooner
than for the more compressible EOS and moderates the differences
in the final spectator momenta.  Figure~\ref{eff} shows
differences in
the Landau effective mass, $m^* = p/v$, in cold nuclear matter
at different densities for MI and MD MFs. Lower masses for the MD
MF means that particles move out faster at same momenta.

\begin{figure}
\centerline{\includegraphics[angle=180,
width=.70\linewidth]{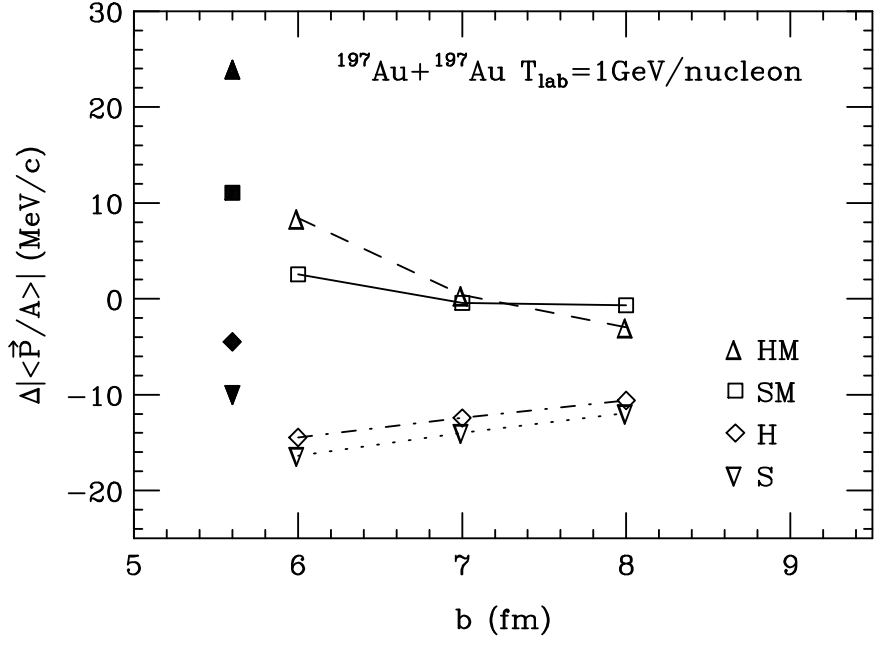}} 
\vspace*{.3in}
\caption{
The change in the net average c.m.\ momentum per nucleon
$\Delta |\langle {\bf P}/A \rangle|$
of the spectator in the $^{197}$Au + $^{197}$Au system
at $T_{lab}=1$~GeV/nucleon.
Open symbols represents results obtained with reduced in-medium
nucleon-nucleon cross sections; filled symbols represent results
obtained at $b=6$~fm with free cross sections.
A negative value of $\Delta |\langle {\bf P}/A \rangle|$
indicates a spectator deceleration,
while a positive value indicates a net acceleration.
}
\label{auau}
\end{figure}

The change in the
magnitude of the c.m.\
momentum per nucleon $\Delta | \langle {\bf P} /A \rangle | $
is generally
dominated by the change in the
longitudinal
momentum per nucleon.  In Figs.~\ref{pxb} and \ref{pxe} the net
momentum per nucleon is seen to decrease
in the Sn + Sn reactions under all conditions.
That change in the momentum might be
considered a measure of the friction involved in the
interaction of
the spectator with the participant zone.  The friction is due
to mentioned random changes of momenta in collisions between
participants and spectators that,
besides knocking particles off spectators, over time drive the
average momentum towards the system average
of zero.  When examining the
net spectator momentum per nucleon as a function of
time in
the Sn + Sn reactions, the momentum is first found to decrease
but then found to recover somewhat.
The late increase and part of the early momentum decrease could
partly be attributed
to our inability of cleanly separating the spectators from the
participants, other than by a convention; the particles are
intermittedly intermixed and then separate.  The above view on
the net spectator momentum, however,
needs to be revised once the changes in the
momentum are examined in the Au + Au system.  The
change in the net momentum per nucleon is shown for 1
GeV/nucleon reaction
as a function of the impact parameter in Fig.~\ref{auau}, by
open symbols for the in-medium reduced cross section.  For low
impact parameters and MD MFs, the
average
spectator momentum per nucleon increases in the reaction
simulations!

The speeding up of the spectator at low $b$ in Au + Au may be,
again, understood in terms of the
explosion of the participant zone.  On one hand, the spectator
acquires the transverse momentum.  On the other, in the
longitudinal direction the explosion acts more on the rear 
of the spectator piece than on
the front.  If the explosion is strong
enough, the ordered push may overcome the friction effects,
producing a net longitudinal acceleration for a piece.  There
is no issue of energy conservation since the work is done by
the participant on the spectator zone.
The difference between Sn + Sn and Au + Au is
in the equilibration time scale relative to the
duration of the fireball.  Differences in the
net final momentum per nucleon between different MFs
for
both systems, with significantly higher net momenta for the MD
than MI MFs, may be understood in terms of the violence of the 
explosion that accelerates the spectator.

An important aspect of the
spectator momentum per nucleon, underscoring the interpretation
above, is its dependence on
the nucleon-nucleon cross section.
In Fig.~\ref{auau}, the results of $b=6$~fm Au+Au simulation
with the free cross sections are represented by filled symbols.
With the larger free cross sections,
the spectator remnants emerge even faster from the reaction than
the with the lower cross sections!
This is because for higher
cross sections, the equilibration is faster,
which allows the participant to explode more violently when
the spectators are still nearby.  Quantitatively, in the
$b=6$~fm HM free cross-section case,
the gain in the longitudinal momentum per nucleon contributes
as much as 17 out of 24 MeV/c of the
gain in the net spectator momentum per nucleon in
Fig.~\ref{auau}.  In the $b=6$~fm HM reduced cross-section
case, the longitudinal gain contributes about 4 out of 8 MeV/c
of the net momentum gain per nucleon.

The mass and the excitation energy of the spectator in
Figs.~\ref{pxb} and \ref{pxe}
do not exhibit a sensitivity to the EOS likely because they are
determined by the geometry and the capability of matter to
retain the energy, respectively.  As to the momentum changes,
though, we have demonstrated that they can provide information
on the violent stage of energetic reactions and constrain
the properties of high density nuclear matter.

\section{Summary and conclusion}
\label{sec:summary}

Within semiclassical transport simulations of energetic
semicentral collisions of heavy ions,
we have carried out an investigation of the interplay between
the participant and spectator regions.
The spectators pass by
the participant region when the participant matter undergoes a
violent explosion.  On one hand, the spectators block the
expansion of the participant matter in the in-plane direction
producing the elliptic
flow for the participant matter.  On
the other hand, the explosion pushes the spectators giving
them transverse momentum pointed away from the reaction
zone.  The momentum transfer to the spectators and the shadow
left in the pattern of the participant emission depend on the
speed of the explosion.  The speed, in turn, depends on the EOS
of the dense matter.  Due to their nature, the spectators
represent a perfectly timed probe right at the reaction site.
A careful analysis of in-plane transverse momentum of a
spectator may yield information on the EOS comparable to that
provided by elliptic flow analysis.  An analysis of the
longitudinal momentum transfer may yield information on the
momentum dependence of the MFs in the reactions.  The
signatures in the spectator momenta per nucleon rise with the
lowering of the impact parameter, but at the cost of the
lowering of a spectator mass, reducing the chances of
identifying the spectator remnants.  Significantly, for most
repulsive MFs and low impact parameters in a heavy system,
spectators may emerge from the reaction with a higher
net average momentum per nucleon than the original momentum.

\acknowledgements

The authors would like to thank K.-H.\ Schmidt and B.\ Tsang
for discussions.
This work was partially supported by the National Science
Foundation under Grant PHY-0070818 and by the Department of
Energy under Grant DE-FGO2-87ER40331.A008.

\newpage
\begin{table}
\caption{Parameter values for the different mean fields
utilized in the simulations. First three columns refer to
Eq.~(\protect\ref{eq:4}) and the next two to
Eq.~(\protect\ref{eq:6}).  The last column gives the Landau
effective mass in normal matter at Fermi momentum.}

\begin{center}
\begin{tabular}{cccccccc} {EOS} & $a$ & $b$ & $\nu$ & $c$ &
$\lambda$ & $m^*/m$ \\
{} & (MeV) & (MeV) & &  & & & \\
\hline
S & 187.24 & 102.623 & 1.6340 &  &  & 0.98  \\
SM & 209.79 & 69.757 & 1.4623 & 0.64570 & 0.95460 & 0.70 \\
H & 121.258 & 52.102 & 2.4624 &  &  & 0.98 & \\
HM &122.785 & 20.427 & 2.7059 & 0.64570 & 0.95460 & 0.70 \\
\end{tabular}
\end{center}
\label{tabmo}
\end{table}

\newpage

\begin{figure}
\centerline{\includegraphics[angle=180,
width=.70\linewidth]{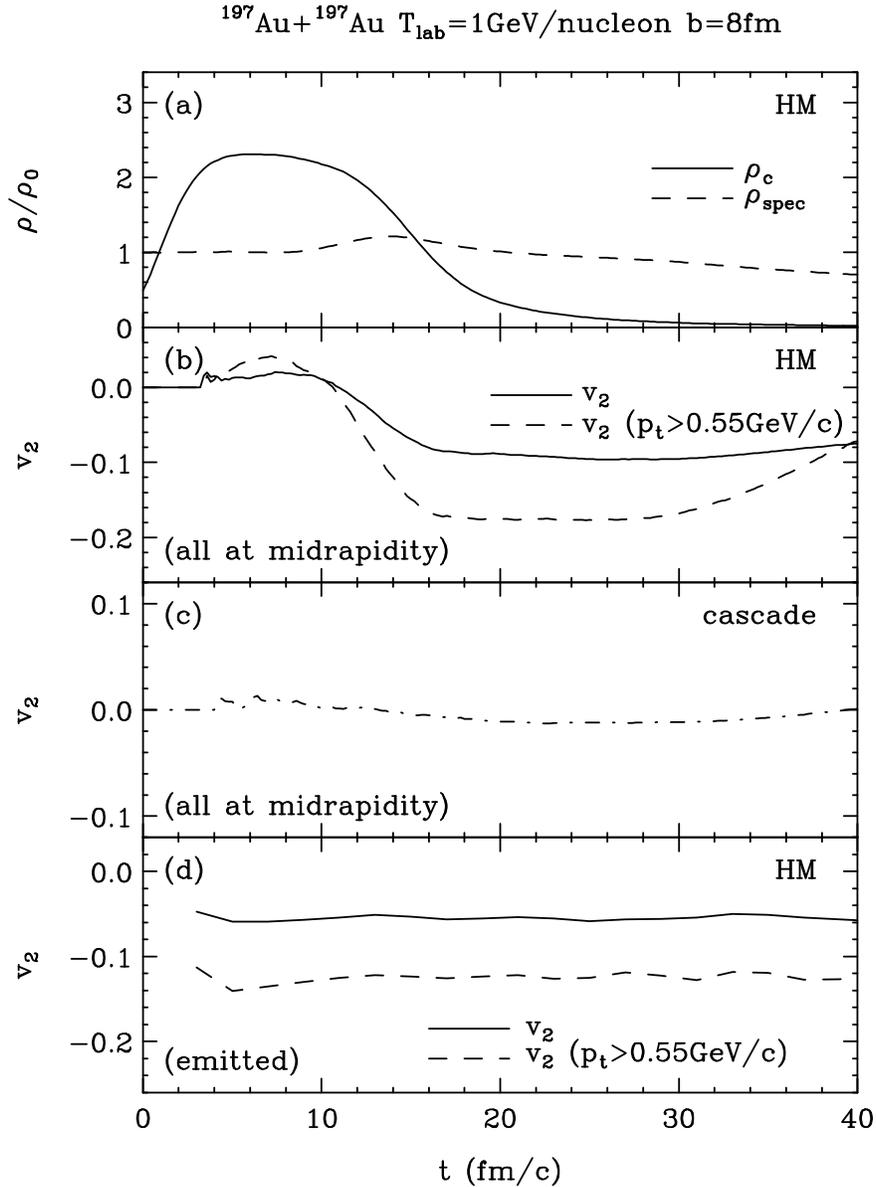}} 
\vspace*{.3in}
\caption{Results from a BUU simulation of the
the $^{197}$Au + $^{197}$Au collision at 1~GeV/nucleon and
$b=8$~fm, as a function of time:
(a) the central densities of the participant~$\rho_c$ and
the spectator  matter~$\rho_{spec}$, (b-d) the
midrapidity elliptical flow
parameter $v_2$.
The results are from a simulation with the HM mean field,
except for those in the panel
(c) which are from
a simulation with no mean field.  The panels (b) and (c)
show the elliptic flow parameter for all particles in the
system while (d) shows the elliptic flow for particles
emitted in the vicinity of a given time.  In the case of the HM
calculations, also shown is $v_2$ when a
high-momentum gate $p_t > 0.55$~GeV/c is applied to the
particles. }
\label{v2t}
\end{figure}

\newpage

\newpage

\newpage

\newpage

\newpage

\newpage

\newpage

\newpage

\end{document}